\documentstyle[11pt,newpasp,twoside,psfig]{article}
\def\edcomment#1{\iffalse\marginpar{\raggedright\sl#1\/}\else\relax\fi}
\reversemarginpar

\begin{document}
\title{RXTE Observations of the Vela Pulsar: \\The X-ray-Optical Connection}
 \author{Alice K Harding}
\affil{NASA Goddard Space Flight Center, Greenbelt, MD 20771 USA}
\author{Mark S. Strickman}
\affil{Code 7651.2, Naval Research Lab., Washington, DC USA}
\author{Carl Gwinn}
\affil{Dept. of Physics, University of California, Santa Barbara, CA USA}
\author{P. McCulloch and D. Moffet}
\affil{University of Tasmania, Tasmania, Australia}

\begin{abstract}
We report on our analysis of a 300 ks observation of the Vela 
pulsar with the Rossi X-Ray Timing Explorer (RXTE).  
The double-peaked, pulsed emission at 2 - 30 keV, which we had 
previously detected during a 93 ks observation, is confirmed 
with much improved statistics.  There is now clear evidence, both
in the spectrum and the light curve, that the emission in the RXTE
band is a blend of two separate components.  The spectrum of the
harder component connects smoothly with the OSSE, COMPTEL
and EGRET spectrum and the peaks in the light curve are in phase
coincidence with those of the high-energy light curve.  The 
spectrum of the softer component is consistent with an 
extrapolation to the pulsed optical flux, and the second RXTE pulse
is in phase coincidence with the second optical peak.  In addition,
we see a peak in the 2-8 keV RXTE pulse profile at the radio phase.
\end{abstract}

\section{Introduction}

The Vela pulsar (PSR B0833-45) is the strongest $\gamma$-ray source in the 
sky, but it is one of the most difficult pulsars to detect at X-ray energies.
This is in part because it is embedded in a very bright X-ray synchrotron nebula
providing a large unpulsed background, but also because its pulsed X-ray emission 
is comparatively weak.  The first detection of pulsed emission at X-ray energies was 
made by ROSAT in the 0.1 - 2 keV band (Ogelman 1993), and the spectrum
is consistent with a blackbody.  We detected the pulsar for the first 
time in hard X-rays (2 - 30 keV) during a 93 ks RXTE Cycle 1 observation (Strickman,
Harding \& De Jager 1999 [SHD99]) and the pulse profile shows two peaks.  The first
RXTE peak is closely aligned with the first EGRET $\gamma$-ray peak, but the second
peak has an energy-dependent phase and is aligned with the second EGRET peak
only at the highest energy (16 - 30 keV).  In our lowest energy RXTE band (2 - 8 keV)
the second peak is roughly aligned with the second peak of the optical profile
(Gouiffes 1998).  The average pulse spectrum joins smoothly with the high-energy
spectrum of OSSE, COMPTEL and EGRET, although the spectrum of the first peak is
significantly harder than that of the second peak, which appeared to be consistent
with an extrapolation to the optical flux points.  We (SHD99) suggested that the
second RXTE peak was a blend of separate hard and soft components.  In this 
paper, we report preliminary results of our analysis of a 300 ks RXTE Cycle 3  
observation which confirms this picture.  

\section{Cycle 3 RXTE Observations and Analysis}

\begin{figure} % Fig. 1
%\vskip 8.0 truecm
\centerline{\psfig{file=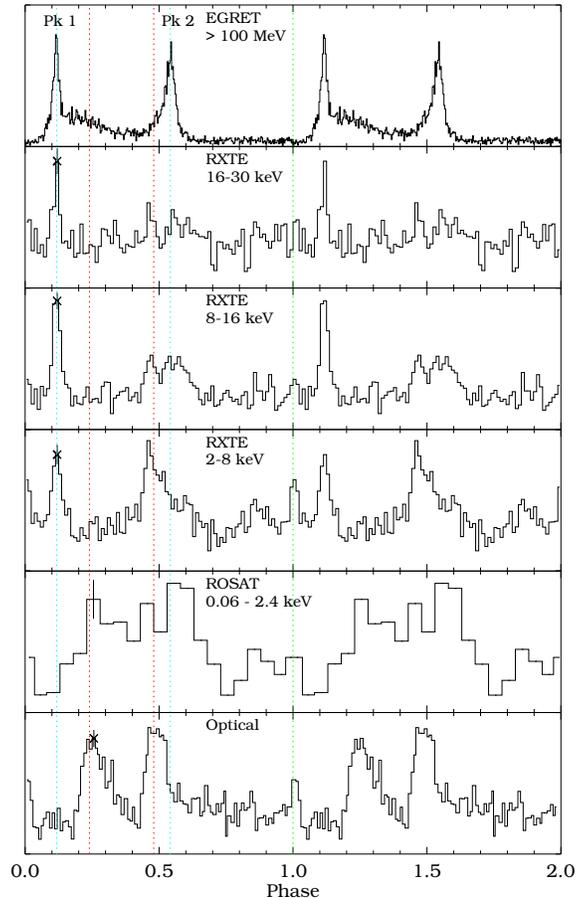,height=13.5cm}}
\vskip -0.7 truecm
\caption[]{Light curves of Cycle 3 RXTE pulsed emission in three bands, shown together
with light curves in EGRET (Kanbach et al. 1994), ROSAT
(Ogelman 1993) and optical (Gouiffes 1998) bands.}
\end{figure}

The Cycle 3 observations were carried out during April/May and July/August 1999 with a
good exposure of 274 ks.  For this analysis we epoch-folded the PCA data in GoodXenon 
event-by-event mode at the pulsar period using the Princeton Pulsar Database
(Arzoumian et al. 1992), to obtain the energy-dependent light curves which we
have summed into three broad energy bands shown in Figure 1.
To separate the various possible components of the light curve and compute individual
component spectra, we fit a five peak
sinusoid model with peaks of the form
\begin{eqnarray}
C(i) & = & A(i) |\cos[{\pi\over 2}(\phi - \phi_{_0}(i))]|^{\xi(i)}, \\
\nonumber \\
\xi(i) & = & -{0.693\over \log[\cos({\pi\over 2}W(i))]} \nonumber
\end{eqnarray}
to the data in 94 energy channels (roughly spanning 2 - 30 keV), where $A(i)$, 
$\phi_0(i)$ and $W(i)$ are the amplitude, center phase and width of peak $i$
respectively.  The model also includes a constant background level.  
We found that $\phi_0(i)$ and $W(i)$ did not vary significantly with energy,
except for the phase of Peak 3 which has a modest variation (see Table 1),
so they were fixed while the
values of $A(i)$ as a function of energy were determined. The counts for each
peak were then found by integrating each sinusoid curve over phase.  We could
then obtain the photon spectrum of each peak separately by fitting a power law with
photoelectric absorption of fixed column density $10^{20}\,\rm cm^{-2}$.

\begin{table}
\caption{Center Phase and Spectrum of Light Curve Peaks} \label{tbl-1}
\begin{center}
\begin{tabular}{lllr}
\tableline
Peak & $\phi_0$ & Photon index & $\chi^2$ (92 dof) \\
\tableline
1 & $0.117 \pm 0.001$ & $0.8892\pm 0.0923$ & 92.8\\
2 & $0.463 \pm 0.006$ & $1.848 \pm 0.244$ & 132.6\\
3 & $0.55  \pm 0.008$ (0.50-0.6)& $1.462 \pm 0.132$ & 99.8\\
4 & $0.87 \pm 0.02$ & $1.463 \pm 0.445$ & 119.9\\
5 & $1.006 \pm 0.004$ & $2.057 \pm 0.312$ & 127.9\\
\hline
\end{tabular}
\end{center}
\vskip -1.0 truecm
\end{table}

\section{Results}

The RXTE light curves in Figure 1 show a narrow peak (Peak 1) at the phase of the EGRET
first peak and a second peak that is now clearly seen to be a blend of two
components.  The first of these components, which we call Peak 2, becomes dominant
in the 2 - 8 keV band and is in phase with the optical second peak.  The second
component, Peak 3, is harder and is in phase with the EGRET second peak.  There are
two other statistically significant peaks that appear in the RXTE light curve:
a peak (Peak 5) at the radio phase (0.0) and a weaker peak (Peak4) 
leading the radio peak.
Note that there is also a peak in the optical light curve at the radio phase.
The energy spectra have been computed using the sinusoid model fits to each peak.
The resulting power law spectral indices and goodness of fit for each peak
are listed in Table 1.  
The peaks in phase with the EGRET peaks (Peak 1 and Peak 3) both have hard, but
significantly different, spectra.  The spectra of Peak 2 and Peak 5 are much softer
and both their flux levels and indices are consistent within the uncertainties.
Interestingly, an extrapolation of the Peak 2 spectrum falls near the optical 
flux points (Nasuti et al. 1997). 

\section{Discussion}

These Cycle 3 observations have thus independently confirmed the multicomponent 
nature of pulsed emission from the Vela pulsar in the energy range 2 - 30 keV suggested 
by the Cycle 1 observations.  With the improved statistics of the Cycle 3 data, we 
have been able to separate the broad second peak in the RXTE X-ray light curve into 
soft (Peak 2) and hard (Peak3) spectral components which maintain their phase integrity throughout the RXTE energy range.  In addition, we have discovered a new feature in
the RXTE light curve: a peak (Peak 5) at the phase of the radio pulse with an extremely 
soft spectrum.  There is, in addition, significant emission leading the radio phase 
(Peak 4).  Peaks 1 and 3 make up the hard spectral component 
whose light curve peaks are in phase with those of the gamma-ray 
light curve, and whose spectrum smoothly connects to the 100 keV - 5 GeV spectrum.
Peaks 2 and 5, whose phases match those of the second optical and radio pulses 
respectively,
make up the soft component.  Their spectra, consistent with each other in both flux
and spectral index, extrapolate to the optical flux points.

Although the RXTE hard component spectrum connects to the $\gamma$-ray spectrum, 
the X-ray spectrum is harder, requiring a break around 100 keV.  Such a break at
the local cyclotron energy, blueshifted by the parallel momentum of the pairs, 
is predicted by the polar cap cascade model (Harding \& 
Daugherty 1999).  The RXTE soft component may be either inverse Compton scattering
radiation of pairs in polar cap cascades (Zhang \& Harding 1999) or synchrotron
radiation of backflowing particles from the outer gap (Cheng \& Zhang 1999).

\end{document}